\documentstyle[12pt]{article}
\begin{document}
\pagestyle{empty}
\begin{titlepage}
\today
\begin{center}
\vspace*{-2cm}
\hfill FTUV/95-66  \\
\hfill IFIC/95-69   \\
\hfill DOE/ER/40561-225-INT95-17-10 \\
\vspace{1cm}
{\Large \bf $\gamma$ - Z interferometry at a $\Phi$-factory } \\
\vspace{1cm}

{\large J. Bernab\'eu}$^a${\large, F.J. Botella}$^{a,b}${\large, O. Vives}$^a$
\\
\vspace*{0.3cm}
$^a$Departament de F\'{\i}sica Te\`orica, Universitat de Val\`encia and
I.F.I.C. (Centre Mixte Universitat de Val\`encia--C.S.I.C.) \\
E-46100 Burjassot (Val\`encia, Spain)
\\
\vspace*{0.2cm}
$^b$Institute for Nuclear Theory, University of Washington \\
Seattle. WA 98195
\vspace*{0.5cm} \\
\begin{abstract}
We analyze the possibilities that the proposed $\Phi$-factories offer
to measure $\gamma-Z$ interference. In the unpolarized beam case, we study
different signatures in the $\rho \pi$ channel, taking advantage of the
presence of the near-by $a_1$ resonance. We build a C-odd forward-backward
asymmetry, estimated to be around $10^{-5}$, and (P-even, T-even) and
(P-odd, T-odd) alignments of the $\rho$, to be seen from the angular
distribution of its $\pi \pi$ decay products. With polarized electrons a
left-right asymmetry around $2\times 10^{-4}$ is present in all channels.
At leading order this asymmetry is independent of hadronic matrix elements
and is sensitive to the $Z^0-s\bar{s}$ vector coupling. In the $\rho \pi$
channel, a combined left-right forward-backward asymmetry is considered.
\end{abstract}
\end{center}
\end{titlepage}
\newpage
\pagestyle{plain}
\pagenumbering{arabic}
\vspace{4cm}

\section{Introduction}
A $\Phi$-factory is an $e^+e^-$ collider operating at a center of
mass energy $q^2 \simeq (m_\Phi)^2 = (1.01941\, GeV)^2$, so that the dominant
process is the formation of the vector resonance $\Phi(1020)$ with spin,
parity and charge conjugation $J^{PC} = 1^{--}$. Projected luminosities of
$10^{32}-10^{33}\, cm^{-2}s^{-1}$ at DA$\Phi$NE and Novosibirsk \cite{fac1},
would produce more than $10^{10}\ \Phi$'s per year, a high number that will
allow to performe high precision experiments. Different possibilities of
testing CP, CPT, chiral lagrangians, etc, can be found in the first and
second editions of the DA$\Phi$NE Physics Handbook  \cite{dafne}.

In this paper we are interested in analyzing the possibilities that a
$\Phi$-factory offers to study $\gamma-Z$ interference effects. That is,
interference among the $\gamma$-dominant contribution to a given final state
F (as indicated in Figure~1) and the contribution of the Z-boson
(see for example Figure 2).

The interference of Figures 1 and 2 is controlled by the ratio of the
Z to $\gamma$ propagators, so its order of magnitude is $q^2/M_Z^2
\simeq 10 ^{-4}$. With $10^{10}$ events per year it seems that this
interference could be measured at a $\Phi$-factory like DA$\Phi$NE.
{}From this kind of measurement one could isolate the $Z-s\bar{s}$
coupling, in fact the vector coupling $(g_V^s)$, because the
$\Phi$ is a pure $s\bar{s}$ vector state. Note that the $Z$ also couples
to the axial current so there is a contribution coming from the nearest-by
axial vector resonance
($J^{PC} = 1^{++}$), the $a_1(1260)$. This resonance has a large width
$\Gamma_{a_1}\simeq 400\, MeV$, so at $q^2\simeq(m_\Phi)^2$ the coupling to
the $a_1(1260)$ is still big enough to produce important effects. This
interference of Figure 1 with Figures 3 and 4
could give complementary information on the $a_1$ parameters \cite{phi.form}.
In fact the $a_1$ contribution in Figure 3 is related to
$\tau^-\rightarrow a_1^- \nu_\tau$ by an isospin rotation.

Other near-by axial vector resonance with a large width is the $h_1(1170)$,
a $J^{PC}=1^{+-}$ state, but the $\gamma$ and the Z can
not couple to this state
through the fist-class vector or axial-vector currents because it is a CP odd
state.
The question of CP violating interferences will be addressed in a
forthcoming paper.

Through this paper we will consider only the interference of Figure 1 with
the Z-mediated ``resonant'' contributions of Figures 2, 3 and 4.
Non-resonant contributions will be neglected, essentially,
because we will only study hadronic final states.

The next point to clarify is where to look for these interference effects.
That is, we have to choose the best channels and the best observables to search
for these interferences. First of all we have to keep in mind that the only
way to see these effects is to look for distinctive signatures. In a
$\gamma -Z$ interference, clearly we only have two signatures of this type.
First, the interference of a P-even amplitude with a P-odd one.
Second, we can have the interference of a C-odd amplitude with a C-even one.
Of course the only kind of signal that unambiguously points towards a
$\gamma- Z$ interference is the first one, that is, to look for parity
violating observables. Nevertheless we will also study C-odd interferences,
but remembering that two photon intermediate states also contribute to
these pieces.

The dominant decay channels of the $\Phi$ are presented in Table I.
Taking into account the number of events necessary to measure an asymmetry of
$10^{-4}$, we will concentrate in the first three channels. In addition we
will distinguish two different cases : we can consider either unpolarized or
polarized beams. DA$\Phi$NE, for example, is going to operate with unpolarized
beams, although electron beam polarization is a possibility that can be
considered \cite{polar}. So we devote the main part of the paper to
study the unpolarized case and at the end we will consider the polarized case,
too.

First we analyze the limited possibilities offered by the $K^+K^-$ and
$K_L K_S$ channels, and then we will study carefully the more promising
$\rho \pi$ channel. We will construct the $\rho$
density matrix and present the observables in the $e^+e^-\rightarrow \rho \pi
\rightarrow \pi \pi \pi$ channels. Again both channels will be analyzed
in the polarized beam case.

\section{$e^-e^+ \rightarrow K^+K^-, K_SK_L$}

The $a_1$ does not couple to $K \bar{K}$ strongly because of Parity. So
in this channel we only have the interference of Figure 1 with Figure 2.
The relevant piece would be when the Z couples to $e^-e^+$ through the
axial current, so we need to construct a pseudoscalar quantity.
But kaons have no spin, and with unpolarized beams we don't have any
pseudoscalar to our disposal in this process.

In conclusion the dominant channels are not useful to measure the  $\gamma- Z$
interference in the case of unpolarized beams. As we will see later, the
situation changes in the case of polarized beams.

\section{$ e^-e^+ \rightarrow \rho \pi$}

In this channel we have a particle with spin in the final state and both
$\Phi$ and $a_1$ couple to $\rho \pi$. So the arguments given in the previous
section go into the opposite direction, that is, we can expect this channel
to be the one where a $\gamma-Z$ interference signal at a
$\Phi$-factory with unpolarized beams can be expected.

\subsection{$ {\bf\rho}$-density matrix ${\bf\rho^{out}}$}

In the process $e^-(l_-,\lambda_-)+e^+(l_+,\lambda_+) \rightarrow
\rho (k,\sigma) + \pi (p)$ for unpolarized initial $e^-e^+$ all the
observables are in the $\rho$-density matrix $\rho^{out}$, because the
pion has no spin.

In the center of mass frame (C.M.), in the so-called helicity convention
\cite{martin}, for an electron with momentum $l_-$ and helicity $\lambda_-$
and a positron with momentum $l_+$ and helicity $\lambda_+$, the amplitude
to a $\pi$ with momentum p, and a $\rho$ with momentum k and helicity
$\sigma$ is given by \cite{martin}

\begin{equation}
\label{ampl 1}
f_{\sigma,\vec{\lambda}}(\theta,\phi) = N \,exp(i\lambda\phi)\,\sum_{J}
\,(2J+1)\, d^J_{\lambda,\sigma}(\theta)\, T^J_{\sigma, \vec{\lambda}}
\end{equation}
where $\vec{\lambda} = (\lambda_-,\lambda_+)$, $\lambda = \lambda_- -
\lambda_+$, $(\theta,\phi)$ are defined in Figure 5, $d^J_{\lambda,\sigma}
(\theta)$ are the reduced rotation matrices
about the y-axis, and the $T^J_{\sigma, \vec{\lambda}}$ are the so-called
reduced helicity amplitudes. In our case, because in all diagrams
$\,e^+e^-$ couple to a vector boson, the sum in Equation (\ref{ampl 1}) gets
reduced to the term $J = 1$. So we can write Equation~(\ref{ampl 1}) in a more
simple way

\begin{equation}
\label{ampl 2}
f_{\sigma,\vec{\lambda}}(\theta,\phi) = \,d^1_{\lambda,\sigma}(\theta)
\, T_{\sigma, \lambda}
\end{equation}

To write Equation (\ref{ampl 2}), we have chosen the scattering plane as
the x-z plane
defined in Figure 5. We have defined $3 N T^1_{\sigma, \vec{\lambda}} =
T_{\sigma, \lambda}$. Finally we have performed the approximation
of neglecting the electron mass; in this case, the vector and axial vector
currents conserve
helicity at the electron vertex. So the only amplitudes we can have are the
ones with $\vec{\lambda} = (+\frac{1}{2},
-\frac{1}{2})$ or \\ $\vec{\lambda} = (-\frac{1}{2},+\frac{1}{2})$, that
means that $\lambda$ can take the values $\lambda = \lambda_- - \lambda_+ =
\pm 1$ and $\lambda$ fixes $\vec{\lambda}$.

In Appendix A we give the $\rho^{out}$ density matrix in general for
processes dominated by vector boson contributions, that is, for the amplitudes
given by Equation (\ref{ampl 2}). If we consider the contribution of the
diagrams in Figures 1 to 4 we can write .

\begin{equation}
\label{reson}
T_{\sigma, \lambda}\,=\,T_{\sigma, \lambda}(\Phi_\gamma)\,+\,
T_{\sigma, \lambda}(\Phi_Z)\,+\,T_{\sigma, \lambda}(a_1^V)\,+\,
T_{\sigma, \lambda}(a_1^A)\,+\, T_{\sigma, \lambda}(\Phi-a_1)
\end{equation}
where $T_{\sigma, \lambda}(\Phi_\gamma)$ is the contribution of Figure~1,
$T_{\sigma, \lambda}(\Phi_Z)$ is the parity violating piece in Figure~2,
the parity conserving piece in Figure~2 will not be considered, because
essentially it has the same structure than Figure~1. $T_{\sigma, \lambda}
(a_1^A)$ and $T_{\sigma, \lambda}(a_1^V)$ are respectively the P-conserving
and P-violating pieces of Figure~3. Finally $T_{\sigma, \lambda}(\Phi-a_1)$
is the contribution of Figure~4, that violates parity.

Now it is evident that $T_{\sigma, \lambda}(\Phi_\gamma)$ and
$T_{\sigma, \lambda}(a_1^A)$ are P even and $T_{\sigma, \lambda}(a_1^V)$ ,
$T_{\sigma, \lambda}(\Phi_Z)$ and $T_{\sigma, \lambda}(\Phi-a_1)$ are P~odd
so they verify \cite{martin}

\begin{eqnarray}
\label{parity}
T_{\sigma, \lambda}(\Phi_\gamma,a_1^A) & = & - T_{-\sigma, -\lambda}
(\Phi_\gamma,a_1^A)
\nonumber \\
T_{\sigma, \lambda}(\Phi_Z,a_1^V,\Phi-a_1) & = & + T_{-\sigma, -\lambda}
(\Phi_Z,a_1^V,\Phi-a_1)
\end{eqnarray}

Also from parity conservation in $\Phi \rightarrow \rho \pi$ and
$a_1 \rightarrow \rho \pi$ decays we have

\begin{eqnarray}
\label{fin.parity}
T_{\sigma, \lambda}(\Phi_\gamma,\Phi_Z) & = & - T_{-\sigma, \lambda}
(\Phi_\gamma,\Phi_Z)
\nonumber \\
T_{\sigma, \lambda}(a_1^A,a_1^V,\Phi-a_1) & = & + T_{-\sigma, \lambda}
(a_1^A,a_1^V,\Phi-a_1)
\end{eqnarray}

Note that a $\Phi$ can not decay to a $\rho$ with helicity $\sigma=0$.
{}From Equations (\ref{parity}) and (\ref{fin.parity}) it is possible to write
all
the reduced helicity amplitudes in terms of the following pieces :
$T_{1,1}(\Phi_\gamma)$, $T_{1,1}(\Phi_Z)$, $T_{1,1}(a_1^A)$,
$T_{1,1}(a_1^V)$, $T_{1,1}(\Phi-a_1)$, $T_{0,1}(a_1^A)$, $T_{0,1}(a_1^V)$ and
$T_{0,1}(\Phi-a_1)$.
Here the most important piece is the first one, being the other
seven of order $10^{-4}$ with respect to the first one, at least.

Using the equations in Appendix A we get for the cross section

\begin{equation}
\label{tot.cross}
\sigma = \int (\frac{d \sigma}{d \Omega}) d \Omega = \frac{16\pi}{3}\,|T_{1,1}
(\Phi_\gamma)|^2
\end{equation}

\begin{equation}
\label{dif.cross}
\frac{d \sigma}{d \Omega} = Tr(\rho^{out}) = \frac{3 \sigma}{16\pi}\,\{
(1 + \cos^2\theta)\,+\,4 Re [\frac{T_{1,1}(a_1^A)}{T_{1,1}(\Phi_\gamma)}]
\cos \theta \}
\end{equation}
with $\theta$ the polar angle of the $\rho$ with respect to the electron
beam.

In Equation (\ref{tot.cross}) we
give the integrated cross section. In fact this equation fixes the
normalization of the reduced helicity amplitudes. Equations (\ref{tot.cross})
and (\ref{dif.cross}) are valid under the assumption of using the
electromagnetic amplitude squared and the $\gamma$-Z interference.
As we can see in Equation (\ref{dif.cross}), it appears for the first time a
$\gamma-Z$ interference.
The second term in the right hand side (r.h.s.) of Equation (\ref{dif.cross})
gives a
forward-backward asymmetry as one would expect from an interference of the
vector$\times$vector (V-V) piece in diagram 1 with the axial$\times$axial
(A-A) piece in diagram 3. It must be stressed at this time that, in
general, any interference of the type (V-V)(A-A) can also be produced by
interference of diagram 1 with two photon processes.
Note that the $2\gamma$ contribution is P-even and C-even as the (A-A) piece
in Figure 3. Two photons in the intermediate state can not couple
to any $0^\pm$ resonance because of helicity supresion. Landau-Yang theorem
\cite{landau} forbids the coupling of two real photons to $1^\pm$ states,
so it remains the resonant contribution $ e^+ e^- \rightarrow \gamma \
\gamma^*\rightarrow a_1^0\rightarrow ...$, where the virtual photon
$\gamma^*$ is coupled to $\rho^0$ or $\omega$.

{}From reference \cite{kuhn}, we can obtain a rigurous lower bound
for this contribution provided the amplitudes $a_1^0 \rightarrow \rho^0
\gamma$, $a_1^0 \rightarrow \omega \gamma$, $\rho^0\rightarrow \gamma$
and $\omega \rightarrow \gamma$ are known. These can be borrowed from
the spin-1 chiral Lagrangian of ref. \cite{ximo} based on an extended
Nambu-Jona-Lasinio model of QCD. With these inputs we get for the ratio
of the absorptive part of the $\gamma(\rho^0,\omega)$ amplitude to the
weak amplitude :

\begin{equation}
\label{gamma.ratio}
\frac{|A(a_1\rightarrow e^+ e^-)_{\gamma \gamma^*}|}{|A(a_1\rightarrow
e^+ e^-)_{weak}|}\geq 8.5
\end{equation}

Although this result indicates that the natural order of magnitude of the
$\gamma \ \gamma^*$ contribution to the forward-backward asymmetry is
larger than the weak one, it must be stressed that there is also the
contribution
of the $\gamma \ \gamma^*$ dispersive part. The forward-backward asymmetry
comes from an interference, so one can get some cancellation or enhancement
between the absorptive and dispersive $\gamma\ \gamma^*$ pieces. Thus no
definite
conclusion about electromagnetic or weak dominance can be extracted from
equation (\ref{gamma.ratio}) till a complete analysis of the process
$e^+e^-\rightarrow \gamma \ \gamma^*\rightarrow a_1$ has been done
\cite{twogamma}.

The next set of observables that can be measured in this channel correspond to
the alignments of the $\rho$, that is, the second-rank multipolar parameters
$t_{2,M}$, defined in Appendix A. For these observables we get

\begin{equation}
\label{alin0}
\frac{d \sigma}{d \Omega}\, t_{2,0}(\theta)\,=\,\sqrt{\frac{1}{10}}\,
\frac{d \sigma}{d \Omega}
\end{equation}

\begin{eqnarray}
\label{alin1}
\frac{d \sigma}{d \Omega}\, t_{2,1}(\theta)&=&\sqrt{\frac{3}{5}}\,\sin\theta
\{Re[T_{0,1}(a_1^A)T_{1,1}^*(\Phi_\gamma)]\nonumber \\
&+&i \cos\theta Im[(T_{0,1}(a_1^V)+
T_{0,1}(\Phi-a_1))\ T_{1,1}^*(\Phi_\gamma)]\}
\end{eqnarray}

\begin{eqnarray}
\label{alin2}
\frac{d \sigma}{d \Omega}\, t_{2,2}(\theta)&=&\sqrt{\frac{3}{5}}\,
\frac{\sin^2\theta}{2} \{ -|T_{1,1}(\Phi_\gamma)|^2\nonumber \\
&+& 2 i Im[(T_{1,1}(a_1^V)+
T_{1,1}(\Phi-a_1))\ T_{1,1}^*(\Phi_\gamma)] \}
\end{eqnarray}

{}From Equations (\ref{alin1}) and (\ref{alin2}) we can see that the imaginary
parts
of $t_{2,1}$ and $t_{2,2}$ contain parity violating pieces of the type
(V-V)(V-A), both from Figures 3 and 4, so these terms unambiguously
point to a $\gamma-Z$ interference.
It must be stressed that the amplitudes that interfere in these observables
must be not relatively real, this points out to T odd observables as we will
see. Note that Figure 2 does not contribute to these Parity violating pieces
, so the Z-$s\bar{s}$ coupling will appear only from the
diagram of Figure 4, where there is a contribution proportional to
$g_V^s$.
In Equation (\ref{alin1}) we have an additional P conserving $\gamma-Z$
interference
that can get contribution also from diagrams with two photons.

For completeness we also give the first-rank multipolar parameters $t_{1,M}
(\theta)$, but in a more standard notation of polarization components.
The angular distributions of the polarization along the x', y' and z' axes
defined in Figure 5 are given by

\begin{equation}
\label{polx}
\frac{d \sigma}{d \Omega}\, P_{x'}(\theta)\,=\,-2 Re[(T_{0,1}(a_1^V)+
T_{0,1}(\Phi-a_1))\ T_{1,1}^*(\Phi_\gamma)]\,\sin\theta \cos\theta
\end{equation}

\begin{equation}
\label{poly}
\frac{d \sigma}{d \Omega}\, P_{y'}(\theta)\,=\,-2 Im[T_{0,1}(a_1^A)
T_{1,1}^*(\Phi_\gamma)]\,\sin\theta
\end{equation}

\begin{eqnarray}
\label{polz}
\frac{d \sigma}{d \Omega}\, P_{z'}(\theta)&=&2 Re[(T_{1,1}(a_1^V)+
T_{1,1}(\Phi-a_1))T_{1,1}^*(\Phi_\gamma)]\,(1 + \cos^2\theta) \nonumber \\
&+& 4 Re[T_{1,1}(\Phi_Z) T_{1,1}^*(\Phi_\gamma)]\, \cos\theta
\end{eqnarray}

As is well-known, the only parity-violating polarizations are within the
scattering plane, and correspond to $P_{x'}$ and $P_{z'}$. $P_{y'}$, being a T
odd observable, needs a relative phase among interfering amplitudes. It has to
be pointed
out that these polarization interferences are rather difficult to measure.
Looking at the
$\rho \rightarrow \pi \pi$ channel is useless, because the pions have no spin,
and $\rho$ decays through a parity conserving interaction.The $\rho^+$ decay
channels are badly measured so we do not know if there is some channel
with branching ratio at the level of $10^{-3}$ that can be used to measure
the $\rho$-polarization. In any case if there is such a channel, even a
polarization at the $10^{-3}$ level is going to be difficult to measure.
So we will not consider $\rho$-polarization as being a measurable quantity
at a $\Phi$-factory, at least at the required level of accuracy to see
$\gamma$-Z interference in the $\rho^\pm \pi^\mp$ channels.

In order to measure the $\rho^0$ polarization, there is a channel that
probably could be used: $\rho^0\rightarrow \pi^+ \pi^- \gamma$ whose branching
ratio is of order $10^{-2}$. In any case sensitivities to the level of
$10^{-4}$ seem difficult to obtain. It must be stressed that there is only
the second piece in the r.h.s. of Equation (\ref{polz}) that can contribute
to polarize the $\rho^0$, $a_1\rightarrow \rho^0\pi^0 $ is forbidden by C,
therefore only the interference with Figure 2 can contribute to the $\rho^0
\pi^0$ channel. Being this piece sensitive to the $Z-s\bar{s}$ vector coupling
($g_V^s$), it would be interesting to analyze the sensitivity to $g_V^s$
of a triple correlation in the process $e^+e^-\rightarrow ``\Phi"
\rightarrow \rho^0 \pi^0 \rightarrow (\pi^+ \pi^- \gamma) \pi^0$
\cite{twogamma}.

Except for this last comment, we must conclude that although there are
several pieces sensitive to the $\gamma-Z$ interference in our observables
, it looks hard if not impossible to measure the contribution of $g_V^s$
coming from Figure 2 with unpolarized beams.

\subsection{The $\rho$-density matrix $\rho^{out}$ and the $Z-s\bar{s}$
coupling}
Let us try to understand Equations (\ref{alin0}) to
(\ref{polz}) in a simple way. Equations (\ref{polx}), (\ref{poly}) and
(\ref{polz}) are proportional to $\langle \vec{s} \cdot
(\alpha\vec{k}+\beta\vec{l_-})
\rangle \approx \langle S_{x'}\rangle \, $
,$\langle \vec{s} \cdot (\vec{k}\times\vec{l_-}) \rangle \approx \langle
S_{y'}\rangle \, $ and
$\langle \vec{s} \cdot \vec{k} \rangle \approx \langle S_{z'}\rangle \, $
respectively, so
clearly $S_{x'}$ and $S_{z'}$ are P-odd and T-even, consequently they
are proportional to the ``real'' part of a parity-violating interference.
On the contrary $S_{y'}$ is P-even and T-odd and then it is proportional to
the ``imaginary'' part of a parity-conserving interference. Note also that
$S_{x'}$ and $S_{y'}$ are a combination of the $S_\pm$ operators, so they need
two different amplitudes that differ by one unit in the helicity of the $\rho$.
However Figure 2 does not populate the $\rho$ in the helicity state $\sigma=0$,
so it can only contribute to $P_{z^\prime}$. The other results
follow inmediatelly from the previous arguments.

If we take into account the following relations for the spherical
components of the tensor polarization

\begin{eqnarray}
\label{ali.pol}
T_{2,0} & \propto & (3 S^2_{z'} - \vec{S}^2)   \\
T_{2,1} & \propto & (S_+ S_{z'} + S_{z'} S_+) \propto  (S_{x'} S_{z'} +
S_{z'} S_{x'}) + i (S_{y'} S_{z'} + S_{z'} S_{y'})
\nonumber \\
T_{2,2} & \propto & S_+^2 \propto  S_{x'}^2 - S_{y'}^2 + i (S_{y'} S_{x'} +
S_{x'} S_{y'})
\nonumber
\end{eqnarray}
it is quite easy to understand the result in Equations (\ref{alin0}),
(\ref{alin1})
and (\ref{alin2}). $\langle T_{2,0} \rangle$, being proportional to $S_i^2$,
must be P-even and T-even, thus the forward-backward asymmetry is proportional
to the ``real'' part of a P-conserving interference. $\langle T_{2,1} \rangle$
has two pieces, the real part proportional to $\langle (S_{x'} S_{z'} +
S_{z'} S_{x'}) \rangle\,$ must be P-even and T-even, and so proportional to the
``real'' part of a P-conserving interference, as it is. The imaginary part
of $\langle T_{2,1} \rangle$ is proportional to $\langle (S_{y'} S_{z'} +
S_{z'} S_{y'}) \rangle\,$, and so it is P-odd and T-odd, consequently it gives
the ``imaginary'' part of a parity-violating interference. The reason why
Figure 2 does not contribute to $T_{2,1}$ is again the absence of population
in the $\sigma= 0$ helicity state of the $\rho$. Being $T_{2,1}$ linear with
$S_{y'}$ the contributions to $\langle T_{2,1} \rangle$ are proportional to
the interference of two amplitudes differing by one unit in the $\rho$
-helicity. On the contrary Figure 4 is proportional to $g_V^s$ and it
contributes to this P-violating piece.

Finally from the last of Eqs. (\ref{ali.pol}), the imaginary part of
$\langle T_{2,2} \rangle$ is itself the imaginary part of a P-violating
interference, because $S_{x'} S_{y'}$ is P-odd and T-odd.
Why Figure 2 does not contribute to this piece ?. If we study only
contributions from Figure 1 and 2, our process can be studied as a two step
process : first the formation of a polarized and aligned $\Phi$,
polarized by the
interference of Figures 1 and 2, and aligned (just $t_{2,0}$) by Figure 1
alone, followed by its decay $A \rightarrow B(J=1)\ +\ C$, it can be checked
than in order to obtain $t_{2,2}\neq 0$ for the B-particle (of spin 1)
, it is necessary to have the A-particle aligned. This is why Figure 1
contributes to the real part of $t_{2,2}$ (P-even and T-even) and why
Figure 2 does not contribute to the imaginary part of $t_{2,2}$ (P-odd
and T-odd); Figure 2 does not aligne the $\Phi$ and consequently does not
contribute to $t_{2,2}$. Note that in this argument we have not used the
fact that $T_{\sigma,\lambda}(\Phi_\gamma)$ and $T_{\sigma,\lambda}(\Phi_Z)$
have the same relative phases, that also implies $Im(t_{2,2})=0$ from
Figure 2. Once again this argument does not apply to Figure 4, which also
contributes to Eq. (\ref{ali.pol}).

In conclusion, we understand why $g_V^s$ coming from Figure 2 appears only
in the second piece of $P_{z'}(\theta)$. Therefore, with unpolarized beams
it seems difficult to extract $g^s_V$ from the contribution of Figure 2
with $\Phi$-factory data, although there is still the contribution of
Figure 4, proportional to $g_V^s$ too.

\section{Observables in $e^-e^+\rightarrow \rho^+ \pi^-
\rightarrow (\pi^+ \pi^0)\pi^-$}

To measure the multipolar parameters of $\rho$, one has to look at the
correlated decay of the $\rho$. To be specific we will consider the channel
$\rho^+ \pi^-$ and the subsequent decay $\rho^+ \rightarrow \pi^+ \pi^0$,
that for our purpose has 100\% branching ratio. The amplitude for a $\rho^+$
with third component of spin $\sigma$ to decay to $\pi^+ \pi^0$ ($\pi^+$ in the
direction $\Omega_1$) in its center of mass frame is given by

\begin{equation}
\label{rhodecay}
f_\sigma (\Omega_1) = (\frac{3}{4\pi})^{\frac{1}{2}} {\cal D}^{(1)}_{\sigma,0}
(\phi_1,\theta_1,0) T
\end{equation}

The single important factor here is the rotation matrix ${\cal
D}^{(1)}_{\sigma,0}
(\phi_1,\theta_1,0)$. $\Omega_1$~is the direction of the $\pi^+$ in the
C.M. of the $\rho$, with respect to the axes we have defined in Figure 5
(x',y',z'). T is the unique reduced helicity amplitude present here,
so it will be controlled by the width $\Gamma(\rho^+\rightarrow \pi^+ \pi^0)
= |T|^2$. Having this channel $100 \%$ branching ratio, we can take $|T|^2$
equal to the total width of the $\rho$, and so we get

\begin{equation}
\label{dospasos}
\frac{d^2\sigma}{d\Omega d\Omega_1} = \frac{1}{|T|^2}
\sum_{\sigma,\sigma^\prime} f_{\sigma}
(\Omega_1) \rho^{out}_{\sigma,\sigma^\prime}(\Omega) f_{\sigma^\prime}^*
(\Omega_1)
\end{equation}

This expresion represents the differential cross section for $e^+e^-
\rightarrow
\rho^+ \pi^-$ and the subsequent decay $\rho^+ \rightarrow \pi^+ \pi^0$ in the
$e^+e^-$ C.M. frame and expressed in terms of $\Omega = (\theta,\phi)$ and
$\Omega_1 = (\theta_1,\phi_1)$, the angle of the $\pi^+$ in the $\rho^+$ C.M..
The result of Eq. (\ref{dospasos}) is independent of the dynamics ($T$)
of $\rho$ decay. Note that  $ \rho^{out}_{\sigma,
\sigma^\prime}$ is invariant under a boost along the direction of motion of the
$\rho$. It is now straight-forward to get

\begin{equation}
\frac{d^2\sigma}{d\Omega d\Omega_1} = \frac{d\sigma}{d\Omega}(e^+e^-\rightarrow
\rho^+ \pi^-)\, \frac{1}{4\pi}\, [1 -\sqrt{10} \sum_N {\cal
D}^{(2)}_{N,0}(\phi_1,
\theta_1,0) t_{2,N}(\theta)]
\end{equation}

As we have previously discussed, the multipolar parameters $t_{1,M}$
(the usual polarizations) have disappeared from the final angular distribution,
and only the second-rank multipolar parameters can be measured. So our final
expresion for the $\rho^+ \pi^-$ channel is

\begin{eqnarray}
\label{ang.distr}
&&\frac{d^2\sigma}{d\Omega d\Omega_1} \,=\, \frac{9\sigma(e^+e^-\rightarrow
\rho^+ \pi^-)}{128 \pi^2}\, \{[(1 + \cos^2\theta) \,+\,
4 Re[\frac{T_{1,1}(a^A_1)}{T_{1,1}(\Phi_\gamma)}]\cos \theta ]\sin^2\theta_1
\nonumber \\
&&+\, [Re[\frac{T_{0,1}(a^A_1)}{T_{1,1}(\Phi_\gamma)}]\,\cos \phi_1 \,
+ Im[
\frac{T_{0,1}(a^V_1)\,+\,T_{0,1}(\Phi-a_1)}{T_{1,1}(\Phi_\gamma)}]\cos\theta
\sin \phi_1 \,]4 \sin\theta \sin \theta_1\nonumber \\
&&\cos \theta_1\,+\,
[\cos(2\phi_1) - 2 Im[\frac{T_{1,1}(a^V_1)\,+
\,T_{1,1}(\Phi-a_1)}{T_{1,1}(\Phi_\gamma)}]\sin(2\phi_1) \, ] \sin^2\theta
\sin^2\theta_1\} \nonumber
\\
\end{eqnarray}

As can be seen from this expresion, the $\gamma-Z$ interference  has produced
four new asymmetries, one of them also present at the $\rho^+\pi^-$ level.
Two of them violate parity and the other two are parity conserving. These
asymmetries can be extracted by fitting the real data with Equation
(\ref{ang.distr}) or by
measuring the following observables

\begin{equation}
\langle \cos \theta \rangle = Re [\frac{T_{1,1}(a^A_1)}{T_{1,1}(\Phi_\gamma)}]
\end{equation}

\begin{equation}
\label{coscos}
\langle \cos \theta_1 \cos \phi_1 \rangle = (\frac{3\pi}{16})^2
\,Re [\frac{T_{0,1}(a^A_1)}{T_{1,1}(\Phi_\gamma)}]
\end{equation}

\begin{equation}
\label{coscossin}
\langle \cos \theta \cos \theta_1 \sin \phi_1 \rangle = \frac{1}{4}
(\frac{3\pi}{16})^2 \,Im [\frac{T_{0,1}(a^V_1)\,+\,T_{0,1}(\Phi-a_1)}
{T_{1,1}(\Phi_\gamma)}]
\end{equation}

\begin{equation}
\label{sin2}
\langle \sin 2\phi_1 \rangle = - \frac{1}{2} Im [\frac{T_{1,1}(a^V_1)\,+\,
T_{1,1}(\Phi-a_1)}{T_{1,1}(\Phi_\gamma)}]
\end{equation}

{}From (\ref{Phi.C}) we see that the $\Phi$ coupling does not change when
going from the $\rho^+\pi^-$ to the $\rho^- \pi^+$ channel, but from
(\ref{a1.C})
the $a_1$ coupling changes sign from $\rho^+\pi^-$ to $\rho^- \pi^+$.
So Equation (\ref{ang.distr}) will also be valid for the $\rho^- \pi^+$
channel,
except that all the asymmetries will change sign. From (\ref{a1.Cl}) we know
that
the $a_1$ does not couple to $\rho^0 \pi^0$, so in this channel these four
asymmetries vanish. Although we have derived these results from Isospin,
actually they can also be obtained from C invariance of strong and
electromagnetic interactions, and so this result is more general.
It must be pointed out that, by comparing different charge channels, one
can eliminate some systematic errors, for example, by performing the ratio of
the difference over the sum of the $\rho^+\pi^-$ and $\rho^-\pi^+$ channels.

Equation (\ref{ang.distr}) has been obtained under the assumption of the narrow
width aproximation for the $\rho$, this means that the imaginary parts of
equations (\ref{coscossin}) and (\ref{sin2}) comes from the relative phase
among the $\Phi$ and $a_1$ propagators at the $\Phi$ peak. If the finite width
of the $\rho$ is taken into account, three intermediate resonant $\rho$'s
have to be included in Figures 1 and 2, and $\rho^+$ and $\rho^-$ in Figures
3 and 4. This fact will bring more phases into the game in such a way
that the results in equations (\ref{coscossin}) and (\ref{sin2}) will change.
These complications will be considered numerically in the next section.
Nevertheless, we must stress
that the inclusion of these additional interferences does not change the
argument in section 3.3 about the absence of any contribution of Figure 2 to
the observable (\ref{sin2}). In fact, the parity violating piece of Figures
1 and 2 with three pions in the final state is proportional to an antisymmetric
leptonic tensor contracted with a "symmetric" hadronic tensor, thus this
contribution vanishes. Note that there is only one form factor for the matrix
element of the vector current between three pions and the vacuum, so in
the \underline{three pion channel} this interference is never present,
even if we include finite widths and interference of three intermediate
$\rho$'s.

\section{Numerical results: unpolarized beams}

Now we will present our estimate of the four asymmetries appearing in
Equation (\ref{ang.distr}). Let us for example concentrate in the one that
gives
a forward-backward asymmetry at the $\rho^+ \pi^-$ level.

\begin{equation}
A_{FB} = \frac{\sigma(\cos \theta >0) - \sigma(\cos \theta <0)}{\sigma(\cos
\theta >0) +\sigma(\cos \theta <0)} = \frac{3}{2} \langle \cos \theta \rangle
= \frac{3}{2} Re [\frac{T_{1,1}(a^A_1)}
{T_{1,1}(\Phi)}]
\end{equation}
{}From Equation (\ref{red11.phigamma}) and (\ref{red11.a1A}) we get

\begin{equation}
\label{AFB2}
A_{FB} = - \frac{9}{2\sqrt{2}}\,(\frac{G_F q^2}{4\pi\alpha})\,\frac{|\vec{k}|}
{k^0} \frac{|F_A(q^2)| |A_1(q^2)|}{|F_V(q^2)| |\Phi(q^2)|} Re [e^{iw_1}\,
\frac{P_{a_1}(q^2)}{P_\Phi(q^2)}]
\end{equation}
where we have substituted the couplings of the standard model, and also we have
introduced a potential relative phase among $F_AA_1$ and $F_V\Phi$, more
likely $w = 0$ or $\pi$. Now we need values for all the form factors appearing
in (\ref{AFB2}).

{}From the width $\Phi \rightarrow e^+e^-$ \cite{phi.form}, Equation
(\ref{prod.phi}), we get

\begin{equation}
|F_V(M_\Phi^2)| = 0.242 \, GeV^2
\end{equation}

{}From $\tau \rightarrow \nu_\tau a_1$, that can be roughly extracted from the
three pion channel \cite{a1.form}, we get using (\ref{a1.tau})

\begin{equation}
|F_A(q^2 \simeq M_{a_1}^2)| = 0.198 \, GeV^2
\end{equation}

This result is also in agreement with a recent lattice calculation
\cite{lattice}, taking into account an $\sqrt{2}$ factor of difference in the
normalization. From  \\ $\Phi \rightarrow \rho^+\pi^-$ \cite{phi.form} and
Equation (\ref{form.phi})
we get

\begin{equation}
|\Phi(M_\Phi^2)| = 1.10 \, GeV^{-1}
\end{equation}

Finally, assuming that $a_1$ decays dominantly \cite{phi.form} to $\rho \pi$
and taking a width of $400\, MeV$ we get

\begin{equation}
\label{a0(q)}
|A_0(q^2 \simeq M_{a_1}^2)| \leq 39.5 \, GeV^{-1}
\end{equation}

\begin{equation}
\label{a1(q)}
|A_1(q^2 \simeq M_{a_1}^2)| \leq 30.8 \, GeV^{-1}
\end{equation}

In principle, there is no apparent dynamical reason why $A_0$ should
be much bigger than $A_1$ or viceversa. Just for illustrative purpose, we
can quote than in the model of reference \cite{Flux} the ratio $A_0/A_1$
gets the value $-1.19$. In this particular model, Equation (\ref{a1.anch})
and $\Gamma_{a_1} = 400 \,MeV$ translate into the following values: $|A_0|
\simeq 26.9\,GeV^{-1}$ and $|A_1| \simeq 22.6\, GeV^{-1}$. ARGUS collaboration
has also measured the $A_0/A_1$ ratio ($S/D$ ratio) \cite{taudecay}
and these results agree, within errors, with the calculation of reference
\cite{Flux}. Then, to estimate the asymmetries we will use these values
from ARGUS measurements, but keeping in mind that the results may
slightly differ from our estimates, because the data are not good enough
in the $\Phi$ region. In fact, if only the limits given by Eqs.
(\ref{a0(q)}) and (\ref{a1(q)}) were considered, all observables could
reach values 1.4 higher than our estimated results.

So just using experimental values and assuming no $q^2$-dependence for the
$|A_1(q^2)|$ form factor we get

\begin{equation}
\label{AFB3}
|A_{FB}(q^2 \simeq M_{a_1}^2)| \simeq 1.63\times 10^{-3}\, | Re [e^{iw_1}\,
\frac{P_{a_1}(q^2)}{P_\Phi(q^2)}]|
\end{equation}

A $\Phi$-factory is going to operate essentially at the $\Phi$-peak,
also the $\Phi$ Breit-Wigner structure can not be extrapolated from the
$\Phi$-peak more than a few times the $\Phi$ width $\Gamma_\Phi$, so
we define the variable $\epsilon$ as $q^2=(M_\Phi +\epsilon\Gamma_\Phi)^2$
and we will consider only the range $-2\leq \epsilon \leq 2$ for illustrations.

For pure Breit-Wigner structures as in (\ref{breit}) we get

\begin{equation}
\label{prop}
\frac{P_{a_1}(q^2)}{P_\Phi(q^2)} = \frac{M_\Phi \Gamma_\Phi}{M_{a_1}
\Gamma_{a_1}}\ G(\epsilon)
\end{equation}
where $G(\epsilon)$ is the function
\begin{equation}
\label{gepsilon}
G(\epsilon) =\{\frac{1 + g(\epsilon) f(\epsilon)}{1 +f^2(\epsilon)} + i
\frac{f(\epsilon) - g(\epsilon)}{1 +f^2(\epsilon)} \}
\end{equation}
and the f and g functions are defined as follows

\begin{equation}
g(\epsilon) = 2 \epsilon (1 + \frac{\epsilon\Gamma_\Phi}{2M_\Phi})
\end{equation}

\begin{equation}
f(\epsilon) = -x + \frac{M_\Phi \Gamma_\Phi}{M_{a_1} \Gamma_{a_1}}\,g(\epsilon)
\end{equation}
$x = (M_{a_1}^2 - M_\Phi^2)/(M_{a_1}\Gamma_{a_1})$. For $\Gamma_{a_1}
= 400\, MeV $ $x$ gets the value 0.98. The real part of Equation (\ref{prop})
vanishes for $\epsilon \simeq 1/(2x)$ and the imaginary one equals to
zero at $\epsilon \simeq -x/2$. These cancellations make difficult to give
any precise value for $A_{FB}$, taking into account: the uncertainties in
$\Gamma_{a_1}$, the poor approximation that represents the pure Breit-Wigner
structure for the $a_1$ and the lack of knowledge of $w_1$. Up to
unfortunate cancellations, it is the factor $(M_\Phi \Gamma_\Phi)/(M_{a_1}
\Gamma_{a_1}) \simeq 9.18 \times 10^{-3}$ that sets the scale in Equation
(\ref{prop}).

So a reasonable estimate of Equation (\ref{AFB3}) would be to take $\epsilon
=0$
and $w_1=0$. In that case we get, always for $\Gamma_{a_1} = 400\,MeV$

\begin{equation}
\label{AFB4}
|A_{FB}(q^2 \simeq M_{a_1}^2)|_{w_1=0} \simeq 7.76 \times 10^{-6}
\end{equation}

In Figure 6 we have plotted $G(\epsilon)$ for $\Gamma_{a_1} = 400
MeV$, and, as far as orders of magnitude are concerned, it can be seen
that both the real and the imaginary parts of $G(\epsilon)$ are similar.

Comparing Equations (\ref{red11.a1A}) and (\ref{red01.a1A}), it is evident
that, up to minor kinematical changes, $Re[\frac{T_{0,1}(a_1^A)}
{T_{1,1}(\Phi_\gamma)}]$
is equal to $\langle \cos \theta \rangle$ except for the change $A_1$ by $A_0$,
so we get in the same spirit than Equation (\ref{AFB4})

\begin{equation}
|Re[\frac{T_{0,1}(a_1^A)}{T_{1,1}(\Phi_\gamma)}]|_{w_1=0} \simeq 6.31
\times 10^{-6}
\end{equation}

Using (\ref{a1-phi}), (\ref{red01.a1v}) and (\ref{red11.a1v}), we can give
the parity violating asymmetries in Equations (\ref{coscossin}) and
(\ref{sin2}), for $sin^2 \theta_W = 0.23$

\begin{eqnarray}
|Im [\frac{T_{0,1}(a_1^V) + T_{0,1}(\Phi-a_1)}{T_{1,1}(\Phi_\gamma)}]|_{w_1=0}
\simeq \nonumber \\
5.27 \times 10^{-3} |Im[(g_V - i \frac{9 \eta g_V^s}{\alpha}
B.r.(\Phi \rightarrow e^+ e^-))\ \frac{P_{a_1}}{P_\Phi}]| =
\nonumber \\
5.27 \times 10^{-3} | g_V Im[\frac{P_{a_1}}{P_\Phi}] - 0.38\eta g_V^s
Re[\frac{P_{a_1}}{P_\Phi}]| = \nonumber \\
(4.85 \times 10^{-2} + 1.44 |g_V^s|) 10^{-5} \simeq 2.98 \times 10^{-6}
\end{eqnarray}

\begin{equation}
|Im [\frac{T_{1,1}(a_1^V) + T_{1,1}(\Phi-a_1)}{T_{1,1}(\Phi_\gamma)}]|_{w_1=0}
\simeq (3.96 \times 10^{-2} + 1.17 |g_V^s| ) 10^{-5} \simeq 2.42 \times 10^{-6}
\end{equation}
As we can see, the dominant piece comes from Figure 4, and so it is
proportional to $g_V^s$. Nevertheless these asymmetries are
difficult to be measured at a $\Phi$-factory. For the observables
(\ref{coscossin}) and (\ref{sin2}) these results translate into

\begin{equation}
\label{ccsnum}
\langle  \cos \theta \cos \theta_1 \sin \phi_1 \rangle = 2.58 \times 10^{-7}
\end{equation}

\begin{equation}
\label{s2num}
\langle \sin 2\phi_1 \rangle = 1.21 \times 10^{-6}
\end{equation}

If finite width effects are included ( $\Gamma_\rho = 150 MeV$) by introducing
three intermediate $\rho$'s in Figure 1 and $\rho^+$ and $\rho^-$ in Figures
3 and 4 the results change a little bit. These observables have to be
integrated over some region of the invariant mass of the $\pi^+ \pi^-$
system. If we take this region to be $[(m_\rho-\Gamma_\rho)^2,(m_\rho+
\Gamma_\rho)^2]$, inside the phase space and a pure Breit-Wigner structure
for the $\rho$-propagator we get

\begin{equation}
\label{fwccsnum}
\langle  \cos \theta \cos \theta_1 \sin \phi_1 \rangle = 1.21 \times 10^{-7}
\end{equation}

\begin{equation}
\label{fws2num}
\langle \sin 2\phi_1 \rangle = 1.35 \times 10^{-6}
\end{equation}

As can be seen from these results, the narrow width approximation is quite
good for $\langle \sin 2\phi_1 \rangle$, but for $\langle  \cos \theta
\cos \theta_1 \sin \phi_1 \rangle$ the corrections due to the finite
( and large) width of the $\rho$ become important.

\section{Polarized beams : Left-Right asymmetries}

As previously mentioned, at a $\Phi$-factory like DA$\Phi$NE , in principle
it would be possible to operate the machine with polarized electrons without
a serious drawback in its luminosity \cite{polar}. So we will consider in
this section
the case when electrons are longitudinally polarized, with polarization P,
and positrons are unpolarized. In this situation it is easy to construct a
pseudoscalar quantity, as is the case for the so-called left-right asymmetry

\begin{equation}
\label{ALR}
A_{LR} = \frac{\sigma(P) -\sigma (-P)}{\sigma(P) +\sigma (-P)}
\end{equation}
where $\sigma(P)$ is the integrated cross section for a given channel F when
the initial electron is longitudinally polarized with value P. This asymmetry
will arise from the interference of the diagrams in Figure 1 and 2, for
any decay channel.

Note that if we integrate over the F variables, the hadronic tensor is
symmetric and
the symmetric part of the leptonic tensor is the same both for the leading term
coming from the Figure 1  - P independent -  and the interference with Figure
2  - this piece being proportional to P - . Other diagrams do not contribute
at leading order to this observable. Taking this fact into account,
we can write $\sigma(P)$ as

\begin{equation}
\label{Pcross}
\sigma(P) = \sigma(P=0) [ 1 \ +\ \frac{16}{\sqrt{2}}\ (\frac{G_F q^2}
{4 \pi \alpha})\ (\frac{g_A g_V^s}{Q_s})\ P ]
\end{equation}

Thus we conclude that in
the ratio of Equation (\ref{ALR}), all the hadronic dependence will cancel out
for any decay channel. Then the result will be

\begin{equation}
\label{ALR2}
A_{LR} = - \frac{12}{\sqrt{2}}\,(\frac{G_F q^2}{4\pi \alpha})\, g^s_V\,P
\end{equation}
where $g^s_V$ is the vector coupling of the Z to $s\bar{s}$ and we have
used $g_A = \frac{1}{4}$ and $Q_s = -\frac{1}{3}$ defined in Appendix C.
We have supposed $\Phi$-dominance for all the channels F.

Equation (\ref{ALR2}) is valid for all decay channels and its value for
$q^2\simeq M_\Phi^2$ and $\sin^2 \theta_W = 0.23$ is $1.94 \times 10^{-4} P$.
So from the statistical point of view, it probably could be measured at
DA$\Phi$NE by going to the dominant channels, that is, $K^+ K^-$, $K_S K_L$
and $\rho \pi$, if a polarized beam of electrons is used.

Other observables can be defined for each particular channel. A Left-Right
differential asymmetry (not integrated in $\theta$) can be defined by

\begin{equation}
\label{LRtheta}
A_{LR}(\theta) = \frac{ \frac{d\sigma}{d\Omega}(\theta, P=+1) -
\frac{d\sigma}{d\Omega}(\theta, P=-1)}{\frac{d\sigma}{d\Omega}(\theta, P=+1)
+\frac{d\sigma}{d\Omega}(\theta, P=-1)}
\end{equation}

{}From these observables we can recover Eq. (\ref{ALR}) by integrating
numerator
and denominator separately. For the $\rho \pi$ channel, for example, we get

\begin{equation}
\label{LRrho}
A_{LR}(\theta) = \frac{ 2 Re[\frac{T_{1,1}(\Phi_Z)}{T_{1,1}(\Phi_\gamma)}]
(1 + \cos^2 \theta) +  4 Re[\frac{T_{1,1}(a_1^V) + T_{1,1}(\Phi - a_1)}
{T_{1,1}(\Phi_\gamma)}]
\cos \theta}{( 1 + \cos^2 \theta)}
\end{equation}
that obviously reproduces Eq. (\ref{ALR}) and verifies the relation

\begin{equation}
\label{pol-LR}
A_{LR}(\theta =0) = P_{z'}(\theta=0)
\end{equation}
as can be checked from Eq. (\ref{polz}). This last relation is a direct
consequence of angular momentum conservation and helicity conservation
at the leptonic vertex. From this general relation we have a consistency
check between Eq. (\ref{LRrho}) and Eq. (\ref{polz}).

Comparing Eqs. (\ref{LRrho}) and (\ref{polz}), we can see that the
angular dependences of the differential left-right asymmetry for polarized
beams, and the longitudinal polarization of the $\rho$ for unpolarized beams
are exchanged. In Eq. (\ref{polz}), the term with $\cos \theta$ is due to
Parity violation in the leptonic vertex, that is, it comes from the decay of
a polarized
$\Phi$, whereas the term with $(1 + \cos^2 \theta)$ is due to Parity
violation in $\Phi$ decay or interference with $a_1$-decay. However,
in the case of initial beam polarization, the term with parity violation
in the leptonic vertex induces a left-right asymmetry even in the total
cross section.

{}From Eq. (\ref{LRrho}), one can build a combined left-right forward-backward
asymmetry as

\begin{equation}
\label{ALRFB}
A^{FB}_{LR} = \frac{\sigma_R(\cos \theta > 0) -\sigma_R(\cos \theta < 0)
-\sigma_L(\cos \theta > 0) +\sigma_L(\cos \theta < 0)}{\sigma_R(\cos \theta >
0) +\sigma_R(\cos \theta < 0) +\sigma_L(\cos \theta > 0)
+\sigma_L(\cos \theta < 0)}
\end{equation}
to separate out the two terms in Eq. (\ref{LRrho}). Under the assumptions
for the $a_1$-vertex form
factors discussed in Section 5, we get $A^{FB}_{LR} \simeq 3.15 \times
10^{-6}$.

\section{Conclusions}
A general analysis of the possibilities of measuring $\gamma-Z$ interference at
a $\Phi$-factory has been worked out. For the case of unpolarized beams we
have looked carefully to the $K^+ K^- $, $K_L K_S$ and $\rho \pi$ channels.
The first two channels do not present any distinctive signature of the
interference. The $\rho \pi$ channel presents several signatures of the
$\gamma-Z$ interference through parity violating observables, taking into
account that the Z boson couples through
the axial current to the $a_1$-meson that dominantly decays to $\rho \pi$.
It is rather unfortunate the fact that the interference of Figures 1 and
2 only contributes to the $\rho$ longitudinal polarization,
a difficult observable to be measured. Nevertheless $g_V^s$
enters in other asymmetries through $\rho$ alignments, but the final values
of these observables are rather small. C-odd asymmetries coming from the (V-V)
(A-A) interference are of the order $10^{-5}$, probably also out of DA$\Phi$NE
performances, in addition other non-weak contribution has to be computed
to give a definite prediction for $\langle \cos\theta \rangle$ and
$\langle \cos\theta_1 \cos\phi_1 \rangle$. Potentially interesting two
photon physics could also be present in these observables.

In the case of $e^+ e^-$ collision at $q^2 \simeq M_\Phi^2$ with longitudinally
polarized electrons and unpolarized positrons, we have obtained that in every
channel there is a left-right asymmetry of order $2\times 10^{-4}$, free
from hadronic uncertainties at leading order and sensitive to the vector
coupling of the Z to the quark s. From the statistical point of view, this
asymmetry seems to be reachable by DA$\Phi$NE operating with polarized
electrons. We have also defined a combined left-right forward-backward
asymmetry to extract the (P-odd,T-even) $a_1-\Phi$ interference.
Its value is much smaller than the left-right asymmetry.

\section*{Acknowledgements}

We would like to acknowledge P. Franzini, P. Herzceg, W. Marciano, M. Musolf,
J. Rosner and G. Vignola for interesting discussions and suggestions. One
of us (F.J.B.) is indebted to W. Haxton for the warm hospitality extended
to him at I.N.T. and to the Conselleria d'Educaci\'o i Ciencia of the
Generalitat Valenciana for financial support. The fellowship from the
Conselleria to O.V. is also acknowledged. This work has been supported by
Grant AEN93-0234 of the Spanish CICYT.

\newpage
\section*{APPENDIX A}
\setcounter{equation}{0}
\def\theequation{A.\arabic{equation}}

The $\rho^{out}$ matrix for the $\rho$ can be written, in the case of
unpolarized beams, as

\begin{equation}
\label{desit.matr}
(\rho^{out})_{\sigma,\sigma^\prime} = \sum_{\vec{\lambda}} f_{\sigma,
\vec{\lambda}} (\theta,\phi)\,f^*_{\sigma^\prime,\vec{\lambda}} (\theta,\phi)
= \sum_\lambda d^1_{\lambda,\sigma}(\theta) d^1_{\lambda,\sigma^\prime}(\theta)
\,T_{\sigma, \lambda} T^*_{\sigma^\prime, \lambda}
\end{equation}

In its center of mass frame or in any boosted frame along z', $\rho^{out}$ can
be written as

\begin{equation}
\label{irr.tensor}
\frac{\rho^{out}}{Tr(\rho^{out})} = \frac{1}{2j+1} \sum_{L,M}^{2j} t_{L,M}^*
T_{L,M}
\end{equation}
where the irreducible tensor operators are refered to the x' ,y' ,z' system
defined by the helicity convention in Figure 5. Definition and normalization
of $T_{L,M}$ follows the conventions in reference \cite{martin}, the physical
observables of the $\rho$ will be given by

\begin{equation}
\label{mult.param}
Tr(\rho^{out}) t_{L,M}(\theta) = Tr (\rho^{out} T_{L,M}) =
\sum_{\sigma, \sigma^\prime} (\rho^{out})_{\sigma, \sigma^\prime} C(1 L 1 |
\sigma M \sigma^\prime)
\end{equation}

The multipole parameters $t_{L,M} = (-)^M t_{L,-M}^*$ define completely the
$\rho^{out}$ density matrix.

For completeness, we also give the $\rho$ polarization in terms of the
multipolar parameters

\begin{equation}
P_{x^\prime} = - (t_{1,1} - t_{1,-1}) = -2\ Re[ t_{1,1}]
\end{equation}

\begin{equation}
P_{y^\prime} = i (t_{1,1} + t_{1,-1}) = -2 \ Im[t_{1,1}]
\end{equation}

\begin{equation}
P_{z^\prime} = \sqrt{2}\ t_{1,0}
\end{equation}

\newpage

\section*{APPENDIX B}
\setcounter{equation}{0}
\def\theequation{B.\arabic{equation}}

In what follows, we shall define the form factors needed to performe our
calculation. Also we shall give expressions of related processes where to
measure these form factors.

The vector form factor is defined by

\begin{equation}
\label{phi.form}
\langle \Phi (\vec{q},\omega)\mid \bar{\psi}_s(0) \gamma_\mu \psi_s(0) \mid 0
\rangle = \  F_V(q^2)\ \varepsilon^*(\omega, \vec{q})_{\mu}
\end{equation}
it represents the coupling of the strange quark vector current to a $\Phi$
of momentum q and helicity $\omega$, $\varepsilon^\mu(\omega,\vec{q})$ is the
polarization four-vector of the $\Phi$. With the one particle states
normalized to $(2 E) (2\pi)^3 \delta^3(\vec{q}- \vec{q^\prime})
\delta_{\omega,\omega^\prime}$, a straightforward calculation for the width of
$\Phi$ going to $e^-e^+$ gives the result

\begin{equation}
\label{prod.phi}
\Gamma_{\Phi\rightarrow e^+ e^-} =\frac{4\pi\alpha^2}{27}\,\frac{|F_V
(M^2_\Phi)|^2}{M_\Phi^{3}}
\end{equation}

The most general matrix element of a $\Phi$ with momentum q and helicity
$\omega$, going to a $\rho$ with momentum k and helicity $\sigma$ and a pion of
momentum p is

\begin{equation}
\label{rho.form}
\langle\rho (\vec{k},\sigma)\ \pi (\vec{p})\mid H \mid \Phi
(\vec{q},\omega) \rangle = i \Phi(q^2)
\ \epsilon_{\mu \nu \alpha \beta}\  k^{\mu} \varepsilon(\omega,\vec{q})^{\nu}
q^{\alpha} \xi^* (\sigma,\vec{k})^{\beta}
\end{equation}
where $\xi (\sigma,\vec{k})^{\beta}$ is the polarization four-vector of the
$\rho$. This result depends upon Lorentz invariance, linearity with the
$\rho$ and $\Phi$ wave functions, and parity conservation. We define $\Phi^{
+,-,0}(q^2)$ corresponding to the three channels $\rho^+\pi^-$, $\rho^-\pi^+$
and $\rho^0\pi^0$. By using isospin we get

\begin{equation}
\label{Phi.C}
\Phi(q^2) = \Phi^{+}(q^2) = \Phi^{-}(q^2)  = -\Phi^{0}(q^2)
\end{equation}

{}From (\ref{rho.form}) we obtain

\begin{equation}
\label{form.phi}
\Gamma(\Phi \rightarrow \rho^+ \pi^-) = \frac{|\vec{k}|^3\ |\Phi(M_\Phi^2)|^2}
{12 \pi}
\end{equation}
if we neglect the mass difference between the $\pi^0$ and the $\pi^\pm$.

The axial form factors are defined in the following way

\begin{equation}
\label{axial.form1}
\frac{1}{2}
\langle a_1^0 (\vec{q},\omega)\mid \bar{\psi}_u(0) \gamma_\mu \gamma_5
\psi_u(0) - \bar{\psi}_d(0) \gamma_\mu \gamma_5 \psi_d(0) \mid 0 \rangle =
F_A(q^2)\ \varepsilon^*(\omega,\vec{q})_{\mu}
\end{equation}

\begin{equation}
\label{axial.form2}
\langle a_1^\pm (\vec{q},\omega)\mid \frac{1}{\sqrt{2}} \bar{\psi}_{u,d}(0)
\gamma_\mu \gamma_5 \psi_{d,u}(0) \mid 0 \rangle =
F_A^\pm(q^2) \ \varepsilon^*(\omega,\vec{q})_{\mu}
\end{equation}

The notation is the same as in (\ref{phi.form}). Because the $a_1$ is
isovector, we have the contribution of the isovector axial-vector
current. From isospin we have

\begin{equation}
\label{a1.Is}
F_A(q^2) = F_A^{+}(q^2) = - F_A^{-}(q^2)
\end{equation}

By treating the $a_1$ in the narrow width approximation, one can extract
\cite{a1.form} from $\tau^- \rightarrow \nu_\tau \pi \pi \pi$ an
aproximate experimental data for  $\tau^- \rightarrow \nu_\tau a_1^-$.
For this process we get

\begin{equation}
\label{a1.tau}
\Gamma(\tau\rightarrow \nu_\tau a_1^-) =
\frac{G_F^2}{8\pi}\ |V_{u,d}|^2\ |F_A^-(M_{a_1}^2)|^2 \frac{M_\tau^3}{M_a^2}
\, (1 - \frac{ M_a^2}{M_\tau^2})^2 \ (1 + 2 \frac{M_a^2}{M_\tau^2})
\end{equation}
Here , $V_{u,d}$ is the corresponding element of the Cabibbo-Kobayashi-Maskawa
matrix, and the rest of the notation is self explanatory.

The most general matrix element of an $a_1$ with momentum q and helicity
$\omega$ going to $\rho \pi$ is

\begin{eqnarray}
\label{a1.form}
\langle \rho^i (\vec{k},\sigma)\ \pi^j (\vec{p}) \mid H_S \mid a_1^k
(\vec{q},\omega) \rangle = A_0^{i,j,k}(q^2)\  q^{\mu}
\xi^*(\sigma,\vec{k})_{\mu}\  k^{\nu} \varepsilon (\omega,\vec{q})_{\nu}
\nonumber \\
+ A_1^{i,j,k}(q^2)\ \{\frac{M_{\rho}^2q^2-(k\cdot q)^2}{(k\cdot q)}
\ \varepsilon (\omega,\vec{q})_{\alpha} \xi^*(\sigma,\vec{k})^{\alpha}
 + q^{\mu} \xi^*(\sigma,\vec{k})_{\mu}\  k^{\nu}
\varepsilon (\omega,\vec{q})_{\nu}) \}
\nonumber \\
\end{eqnarray}
Here we have two form factors: $A_1^{i,j,k}$ contributes to the amplitudes
where the $\rho$ is in the +1 or -1 helicity state, and $A_0^{i,j,k}$
to the 0 helicity state of the $\rho$. The general structure (\ref{a1.form})
depends upon the same asumptions we have used in (\ref{rho.form}). The indices
i, j, k indicate the corresponding charges of each channel. From Isospin
we have

\begin{equation}
\label{a1.Cl}
A_\alpha^{i,j,k} = \sqrt{2} C(1 1 1 | I_i I_j I_k) A_\alpha
\end{equation}
in such a way the $a_1^0$ does not decay to $\rho^0 \pi^0$ and

\begin{equation}
\label{a1.C}
A_\alpha = A_\alpha^{+,-,0} = - A_\alpha^{-,+,0} = A_\alpha^{0,-,-} =
- A_\alpha^{-,0,-}
\end{equation}

We must point out that both (\ref{rho.form}) and (\ref{a1.form}) are valid for
$\rho$ and $\pi$ both on shell, otherwise there would be other dependences
with its invariant masses.

The width of the $a_1$ is given by

\begin{equation}
\label{a1.anch}
\Gamma_{a_1^i\rightarrow \rho^j \pi^k} = \frac{\lambda^{5/2}(M_a^2,M_\rho^2
,M_\pi^2)}{768 \pi M_a^5 M_\rho^2}
\{|A_1^{i,j,k}(q^2)|^2\frac{8 M_\rho^2 M_a^2}{(M_a^2+M_\rho^2-M_\pi^2)^2} +
|A_0^{i,j,k}(q^2)|^2\}
\end{equation}

Separate information on $A_0$ and $A_1$ needs additional information about
the $a_1$-decay, that could be extracted through a careful analysis of the
$\tau \rightarrow \nu_\tau \pi \pi \pi$ decay \cite{taudecay}

\newpage
\section*{APPENDIX C}
\setcounter{equation}{0}
\def\theequation{C.\arabic{equation}}

In this section, we present the calculation of the reduced helicity amplitudes
of the $e^+e^- \rightarrow \rho^+ \pi^-$ process. The Kinematics in the C.M.
frame of the $e^+ e^-$ system (or the $\Phi$ system) is defined by

\begin{eqnarray}
\label{kinematics}
l^\mu_- = (E,0,0,|\vec{l}|) & q^\mu = (l_-+l_+)^\mu \\
l^\mu_+ = (E,0,0,-|\vec{l}|) & \xi^\mu(0,\vec{k}) = \frac{1}{M_\rho}
(|\vec{k}|, k^0\sin\theta,0,k^0\cos\theta) \nonumber \\
k^\mu = (k^0,|\vec{k}|\sin\theta,0,|\vec{k}|\cos\theta) &
\xi^\mu(\pm 1,\vec{k}) = \mp \frac{1}{\sqrt{2}}
(0, \cos\theta,\pm i , -\sin\theta) \nonumber
\end{eqnarray}

The amplitudes we need correspond to the diagrams in Figures 1 to 4.
Every vertex has been previously defined in Appendix B.
The $\Phi$ and the $a_1$ propagators will be taken as pure Breit-Wigner
distributions, although, more sophisticated versions can be taken if necessary.
Relative phases have been imposed by the T properties of the currents, in
such a way that relative imaginary pieces come essentially from the
Breit-Wigner structure. The Feynman amplitudes corresponding to diagrams
1 to 3 are

\begin{equation}
\label{phi-gamma}
M^{\Phi_\gamma}_{\sigma,\lambda_+,\lambda_-} (\theta) = i \frac{4 \pi \alpha}
{q^2} Q_s F_V(q^2) \Phi(q^2) P_\Phi(q^2) V_\mu(\lambda_+,\lambda_-)
\epsilon^{\mu \nu \alpha \beta} \xi_\nu^*(\sigma,\vec{k}) k_\alpha q_\beta
\end{equation}

\begin{eqnarray}
\label{phi-Z}
M^{\Phi_Z}_{\sigma,\lambda_+,\lambda_-} (\theta) &=& i \frac{8 G_F}{\sqrt{2}}
g_V^s F_V(q^2) \Phi(q^2) P_\Phi(q^2) [g_V V^\mu(\lambda_+,\lambda_-) + g_A
A^\mu(\lambda_+,\lambda_-)] \nonumber \\
&&\epsilon^{\mu \nu \alpha \beta} \xi_\nu^*(\sigma,\vec{k}) k_\alpha q_\beta
\end{eqnarray}

\begin{eqnarray}
\label{a1.diagr}
M^{a_1}_{\sigma,\lambda_+,\lambda_-} (\theta) = - \frac{8 G_F}{\sqrt{2}}
g_{3A} F_A(q^2) P_{a_1}(q^2) [g_V V^\mu(\lambda_+,\lambda_-) + g_A
A^\mu(\lambda_+,\lambda_-)] \nonumber \\
\{A_1(q^2)\frac{M_\rho^2 q^2 - (k\cdot q)}{(k \cdot q)}
\xi^*_\mu(\sigma,\vec{k}) + [A_1(q^2) + A_0(q^2)](q \cdot \xi^*(\sigma,\vec{k})
)k_\mu \}
\end{eqnarray}
The amplitude corresponding to diagram 4, $M^{\Phi-a_1}_{\sigma,\lambda_+,
\lambda_-}$, can be obtained from $M^{a_1}_{\sigma,\lambda_+,\lambda_-}$
by taking $g_A= 0$, and performing the substitution
\begin{equation}
\label{a1-phi}
g_V \rightarrow -\frac{4 \pi \alpha}{q^2} P_\Phi(q^2) |F_V(q^2)|^2 Q_s g^s_V
\eta
\end{equation}

In this expresions, $Q_s = -\frac{1}{3}$ is the s-quark charge, $g_V = -
\frac{1}{4} + \sin^2\theta_W$ and $g_A = \frac{1}{4}$ are the Z couplings
to leptons, $g_{3A} = -\frac{1}{2}$ is the isovector axial-vector coupling to
light quarks, and $g_V^s = - \frac{1}{4} + \frac{1}{3} \sin^2 \theta_W$ is the
vector coupling of the Z to the s-quarks. Note that the precise normalization
of the currents entering in
the form factors is defined in (\ref{axial.form2}). $P_\Phi(q^2)$ and
$P_{a_1}(q^2)$ are the $\Phi$ and $a_1$ propagators respectively. We will
take the usual Breit-Wigner structure

\begin{equation}
\label{breit}
P(q^2) = \frac{1}{(q^2 - M^2) +i M \Gamma}
\end{equation}
and the matrix element of the leptonic currents are given by

\begin{eqnarray}
\label{lept.curr}
V^\mu(\lambda_-,\lambda_+) = \bar{v}(l_+,\lambda_+) \gamma^\mu u(l_-,\lambda_-)
= 2 E\ \delta_{-\lambda_+,\lambda_-} (0,\lambda_- -\lambda_+,i,0) \nonumber \\
A^\mu(\lambda_-,\lambda_+) = \bar{v}(l~_+,\lambda_+) \gamma^\mu \gamma_5
u(l_-,\lambda_-) = 2 E\ \delta_{-\lambda_+,\lambda_-} (0,1,
i(\lambda_- -\lambda_+),0) \nonumber \\
\end{eqnarray}

It must be stressed that the weak hamiltonian entering in the amplitude
$M^{\Phi-a_1}_{\sigma,\lambda_+,\lambda_-}$ is a pure non-leptonic one, so
there can be additional contributions. Double
Cabibbo suppressed charged current contributions are also present. From
the analysis \cite{savage}, the QCD corrections give rise to the factor
$\eta \simeq 1.5$ in (\ref{a1-phi}). The diagram in Figure 4 has been
recently considered in ref \cite{chiral}, as ``the parity violating piece
of the weak decay $\Phi \rightarrow \rho \pi$'' in the context of Chiral
Perturbation Theory for vector mesons. We will not follow this calculation,
because $a_1$-dominance has not been included, as the experimental data
in $\tau \rightarrow \pi \pi \pi \nu_\tau$ \cite{taudecay} indicate, even
at 1 GeV.

On the $\Phi$-peak the (r.h.s.) of (\ref{a1-phi}) can be written as

\begin{equation}
\label{phi.peak}
-i\ \frac{9\ \eta\ g_V^s}{\alpha} \ B.r.(\Phi \rightarrow e^+ e^-)
\end{equation}
in such a way that for $\sin^2 \theta_W = 0.23$, it is $3.3 \eta$ times
bigger than the (l.h.s.) in (\ref{a1-phi}).

Finally the normalization in Equations (\ref{phi-gamma}), (\ref{phi-Z}) and
(\ref{a1.diagr}) is such that

\begin{equation}
\label{norm}
\sigma(e^+ e^-\rightarrow \rho\ \pi) = \frac{1}{8 q^2}\int \frac{d^3k}{2 k^0}
\frac{d^3p}{2 p^0} \frac{1}{2\pi^2} \sum_{\sigma,\lambda_+,\lambda_-}
|M_{\sigma,\lambda_+,\lambda_-}(\Phi + a_1)|^2
\delta^{(4)} (q - k -p)
\end{equation}

{}From the $\Phi$-diagram we get
\begin{equation}
\sigma(e^+ e^-\rightarrow \rho\ \pi)\vert_{q^2\simeq M_\Phi^2}
\simeq \frac{12 \pi \Gamma(\Phi \rightarrow e^+e^-) \Gamma(\Phi \rightarrow
\rho^+ \pi^-)}{(q^2 -M_\Phi^2)^2 + M_\Phi^2 \Gamma_\Phi^2}
\end{equation}
where the equality holds in the range where the $q^2$-dependence of
$F_V(q^2)$ and $\Phi(q^2)$ can be neglected, a very good aproximation arround
the $\Phi$-peak.

A lengthy but straightforward calculation has allowed to check the relation

\begin{equation}
M_{\sigma,\lambda_+,\lambda_-} (\theta) = K f_{\sigma,\vec{\lambda}}(\theta)
\end{equation}
for both $\Phi$ and $a_1$ contributions. K is a $q^2$-dependent constant that
takes care of the different normalizations in Equation (\ref{norm}) and
(\ref{dif.cross}) and (\ref{tot.cross}).

Finally we get the following results

\begin{equation}
\label{red11.phigamma}
K T_{1,1}(\Phi_\gamma) = - 4 \pi\ \sqrt{2} \alpha Q_s\ |\vec{k}|\ F_V(q^2)
\Phi(q^2) P_\Phi(q^2)
\end{equation}

\begin{equation}
\label{red11.phiZ}
K T_{1,1}(\Phi_Z) = - 8 \ G_F q^2\ g_V^s\ g_A\ |\vec{k}|\ F_V(q^2) \Phi(q^2)
P_\Phi(q^2)
\end{equation}

\begin{equation}
\label{red01.a1v}
K T_{0,1}(a_1^V) = - 8\ G_F q^2\ g_V g_{3A}\ \frac{|\vec{k}|^2}{M_\rho}\
F_A(q^2)\ A_0(q^2) P_{a_1}(q^2)
\end{equation}

\begin{equation}
\label{red11.a1v}
K T_{1,1}(a_1^V) =  8\ G_F q^2\ g_V g_{3A}\ \frac{|\vec{k}|^2}{k^0}\ F_A(q^2)
A_1(q^2) P_{a_1}(q^2)
\end{equation}

\begin{equation}
\label{red01.a1A}
K T_{0,1}(a_1^A) = - 8\ G_F q^2\ g_A g_{3A}\ \frac{|\vec{k}|^2}{M_\rho}\
F_A(q^2)\ A_0(q^2) P_{a_1}(q^2)
\end{equation}

\begin{equation}
\label{red11.a1A}
K T_{1,1}(a_1^A) =  8\ G_F q^2\ g_A g_{3A}\ \frac{|\vec{k}|^2}{k^0}\ F_A(q^2)
A_1(q^2) P_{a_1}(q^2)
\end{equation}

$K T_{0,1}(\Phi-a_1)$ and $K T_{1,1}(\Phi-a_1)$ can be obtained from
(\ref{red01.a1v}) and (\ref{red11.a1v}) respectively just performing the
substitution (\ref{a1-phi}).

The following relations have been checked out explicitly

\begin{equation}
T_{1,1}(\Phi_\gamma) = T_{1,-1}(\Phi_\gamma) =
- T_{-1,-1}(\Phi_\gamma) = - T_{-1,1}(\Phi_\gamma)
\end{equation}

\begin{equation}
T_{1,1}(\Phi_Z) = - T_{1,-1}(\Phi_Z) =
T_{-1,-1}(\Phi_Z) = - T_{-1,1}(\Phi_Z)
\end{equation}

\begin{eqnarray}
T_{0,1}(a_1^V,\Phi-a_1) &=& T_{0,-1}(a_1^V,\Phi-a_1)
\\  T_{1,1}(a_1^V,\Phi-a_1) = T_{1,-1}(a_1^V,\Phi-a_1) &=&
T_{-1,1}(a_1^V,\Phi-a_1) = T_{-1,-1}(a_1^V,\Phi-a_1) \nonumber
\end{eqnarray}

\begin{eqnarray}
T_{0,1}(a_1^A) &=& - T_{0,-1}(a_1^A)
\\  T_{1,1}(a_1^A) = - T_{1,-1}(a_1^A) &=&
T_{-1,1}(a_1^A) = - T_{-1,-1}(a_1^A) \nonumber
\end{eqnarray}

These are equivalent to Equations (\ref{parity}) and (\ref{fin.parity}).



\newpage
\hspace*{-0.6cm}{\Large \bf Table Captions}\\
\\
\\
\\

{\bf Table I}: Dominant $\Phi$ decay channels.

\newpage
\begin{center}
{\bf Table I}
\end{center}

\begin{table}[t,h,b]
\begin{center}
\begin{tabular}{|l|l|}
\hline
\multicolumn{2}{|c|}{$\Phi (1020)\; \; \;\;  I^G(J^{PC})=0^-(1^{--})$}\\\hline
\multicolumn{2}{|c|}{Mass $\;M_\Phi\; = 1019.413 \pm 0.008\; MeV $}\\
\multicolumn{2}{|c|}{Width $\;\Gamma\;= 4.43 \pm 0.06\;MeV $}\\ \hline
Decay modes  &  Fraction $\Gamma_i /\Gamma $  \\ \hline
 $ K^+ K^- $ & $(49.1 \pm 0.9) \% $  \\
 $ K^0_L K^0_S $ & $ (34.3 \pm 0.7) \% $\\
 $\rho\;\pi$ & $ (12.9 \pm 0.7) \% $\\
 $\pi^+\pi^-\pi^0$ & $ (2.5 \pm 0.9) \% $ \\
 $\eta\;\gamma$ & $(1.28 \pm 0.06) \% $\\
 $\pi^0\;\gamma$ & $(1.31 \pm 0.13) \times 10^{-3} $ \\
 $e^+\;e^-$ & $(3.09 \pm 0.07) \times 10^{-4} $ \\
 $\mu^+\;\mu^-$ & $(2.48 \pm 0.34) \times 10^{-4} $ \\ \hline
\end{tabular}
\end{center}
\end{table}

\newpage
\hspace{-0.6cm}{\Large \bf Figure Captions} \\
    \\
    \\
    \\

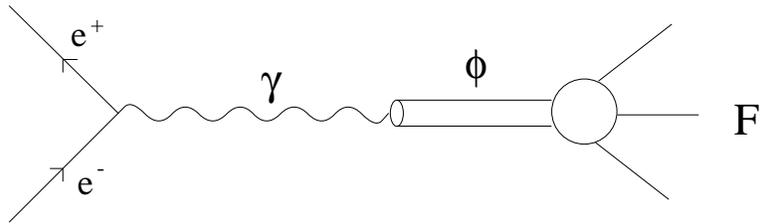
{\bf Figure 1}: Dominat diagram for $\Phi$ production at a $\Phi$-factory.
\\
\\

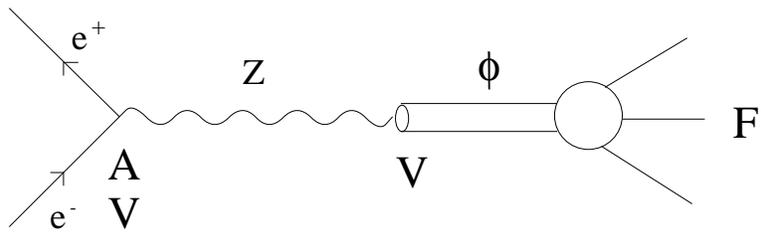
{\bf Figure 2}: Z-mediated production of $\Phi$. V (A) means a vector (axial)
coupling.
\\
\\

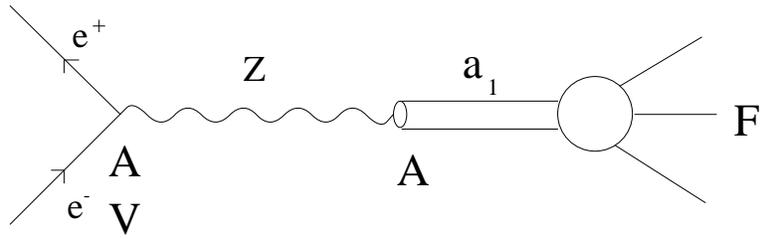
{\bf Figure 3}: Pure $a_1$ resonant contribution, mediated by Z.
\\
\\

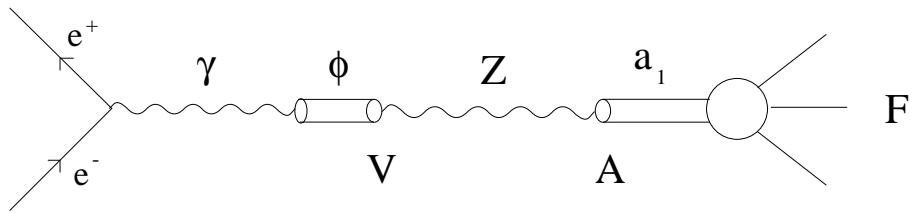
{\bf Figure 4}: Weak decay of $\Phi$ through $a_1$.
\\
\\

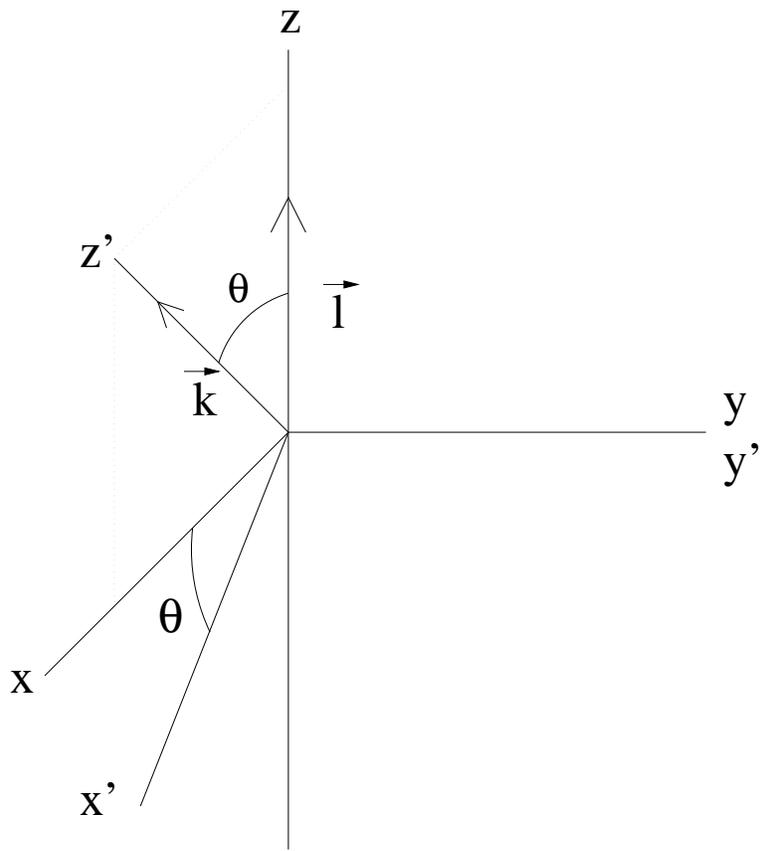
{\bf Figure 5}: Reference system. (x, y, z ) define the axes in the C.M. of the
$e^+ e^-$ system. (x', y', z' ) are the corresponding axes in the $\rho$ C.M.
frame of reference.
\\
\\

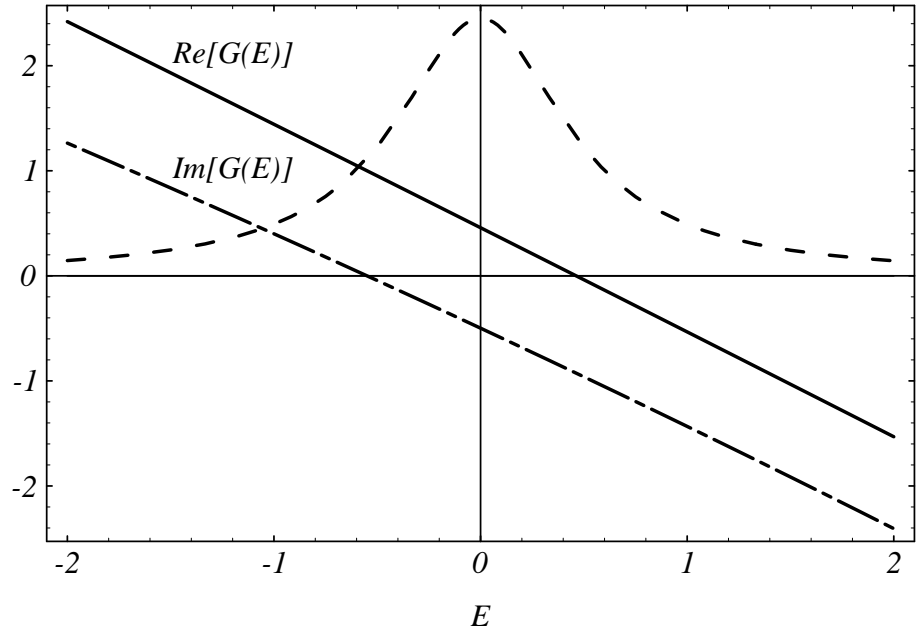
{\bf Figure 6}: Real and Imaginary part of $G(\epsilon)$, defined in Equation
(\ref{gepsilon}). The Breit-Wigner
structure of the $\Phi$ has been included.
\\

\newpage
\begin{thebibliography}{99}

\bibitem{fac1}  G. Vignola, {\em Proceeding of the Worshop on Physics and
		Detectors for DA$\Phi$NE}, ed G. Pancheri, (INFN,
		Frascati 1991), p. 11. \\
		A. N. Skrinsky,{\em Proceeding of the Worshop on Physics and
		Detectors for DA$\Phi$NE}, ed G. Pancheri, (INFN,
		Frascati 1991), p. 67.

\bibitem{dafne} {\em The DA$\Phi$NE Physics Handbook}, eds L. Maiani,
		G. Pancheri and N.Paver, (INFN, Frascati 1992). \\
		{\em The Second DA$\Phi$NE Physics Handbook}, eds L. Maiani,
		G. Pancheri and N.Paver, (INFN, Frascati 1995). \\

\bibitem{phi.form} Particle Data Group, M. Aguilar-Benitez {\em et al.},
		  {\em Phys. Rev.} {\bf D50}, 1173 (1994).

\bibitem{polar} Private comunication by G. Vignola.

\bibitem{martin} A.D. Martin, T.D. Spearman, ``Elementary Particle Theory'',
		 North-Holland, Amsterdam 1970.

\bibitem{landau}  L.D. Landau, {\em Dokl. Akad. Nawk.}, USSR {\bf 60}, 207
		  (1948). Sumary in English in {\em Phys. Abstracts}
		  {\bf A52}, 125 (1949). \\
		  C.N. Yang, {\em Phys. Rev.} {\bf 77}, 242 (1950).

\bibitem{kuhn} J.H. K\"uhn, J. Kaplan and E.G.O. Safiani, {\em Nucl. Phys.}
		{\bf B157}, 125 (1979)

\bibitem{ximo} J. Prades, {\em Z. Phys} {\bf C63}, 491 (1994)

\bibitem{twogamma} Work in progress.

\bibitem{a1.form} W.T. Ford, {\em Nucl. Phys. B (Proc. Suppl.)} {\bf 40},
		  191 (1995). \\
		  J.H. K\"{u}hn and A. Santamaria, {\em Z. Phys.} {\bf C48}
		  , 445 (1990).

\bibitem{lattice} M. Wingate, T. DeGrand, S. Collins, and U. Heller,
		  {\em Phys. Rev. Lett.} {\bf 74}, 4596 (1995).

\bibitem{Flux} N. Isgur,  C. Morningstar and C. Reader, {\em Phys. Rev.}
	       {\bf D39}, 1357 (1989).

\bibitem{taudecay} ARGUS Collaboration, H. Albretch {\em et al.},
	           {\em Z. Phys.} {\bf C58}, 61 (1993).

\bibitem{savage} J. Dai, M.J. Savage, J. Liu and R. Springer, {\em Phys.
  		 Lett.} {\bf B271}, 403 (1991).

\bibitem{chiral} E. Jenkins, A. Manohar, and M. Wise, {\em Phys. Rev. Lett.}
		 {\bf 75}, 2272 (1995).

\end{thebibliography}
\end{document}